# Raoult's law revisited: accurately predicting equilibrium relative humidity points for humidity control experiments.

Michael G. Bowler[a,*], David R. Bowler[b] and Matthew W. Bowler[c,d,*]

a) Department of Physics, University of Oxford, Keble Road, Oxford, OX1 3RH, UK
b) Department of Physics and Astronomy, University College London, Gower St, London, WC1E 6BT, UK
c) European Molecular Biology Laboratory, Grenoble Outstation, 71 avenue des Martyrs, CS 90181 F-38042 Grenoble, France
d) Unit for Virus Host Cell Interactions, Univ. Grenoble Alpes-EMBL-CNRS, 71 avenue des Martyrs, CS 90181 F-38042 Grenoble, France

[*]Corresponding authors:

Michael G. Bowler
Department of Physics,
University of Oxford,
Keble Road,
Oxford OX1 3RH
UK
Email: michael.bowler@physics.ox.ac.uk

Matthew W. Bowler
European Molecular Biology Laboratory,
Grenoble Outstation,
71 avenue des Martyrs,
CS 90181 F-38042 Grenoble
France
Email: mbowler@embl.fr

**Synopsis:** The equilibrium relative humidity values for a number of the most commonly used precipitants in biological macromolecule crystallisation have been measured using a new humidity control device. A simple argument in statistical mechanics demonstrates that the saturated vapour pressure of a solvent is proportional to its mole fraction in an ideal solution (Raoult's Law). The same argument can be extended to the case where solvent and solute molecules are of different size.

**Abstract**

The humidity surrounding a sample is an important variable in scientific experiments. Biological samples in particular require not just a humid atmosphere but often a relative humidity (RH) that is in equilibrium with a stabilizing solution required to maintain the sample in the same state during measurements. The controlled dehydration of macromolecular crystals can lead to significant increases in crystal order, which often leads to higher diffraction quality. Devices that can accurately control the humidity surrounding crystals on a beamline have led to this technique being increasingly adopted, as experiments become easier and more reproducible. Matching the relative humidity to the mother liquor is the first step to allow the stable mounting of a crystal. In previous work, we measured the equilibrium relative humidity for a range of concentrations of the most commonly used precipitants and showed how this related to Raoult's law for the equilibrium vapour pressure of water above a solution. However, a discrepancy between measured values and those predicted by theory could not be explained. Here, we have used a more precise humidity control device to determine equilibrium relative humidity points. The new results are in agreement with Raoult's law. We also present a simple argument in statistical mechanics demonstrating that the saturated vapour pressure of a solvent is proportional to its mole fraction in an ideal solution: Raoult's Law. The same argument can be extended to the case where solvent and solute molecules are of different size, as is the case with polymers. The results provide a framework for the correct maintenance of the RH surrounding samples.

**1.0 Introduction**

Sample environments that control relative humidity (RH) are important in many experiments where a wide variety of samples require specific RH values to maintain sample integrity or RH is a parameter to be varied. Humidity control has been an important parameter in the study of lipid bilayers (Lin *et al.*, 2007), amyloid fibers (McDonald *et al.*, 2008), small molecule crystallography (Mo & Ramsoskar, 2009) as well as coherent X-ray diffraction microscopy of cells (Takayama & Nakasako, 2012). In biological crystallography, changing the RH can often induce phase changes in crystals of macromolecules with the concomitant improvement in the quality of observed diffraction. This has been observed since the earliest days of macromolecular crystallography (Berthou *et al.*, 1972; Einstein & Low, 1962; Huxley & Kendrew, 1953; Perutz, 1946) and is most easily effected by altering the molar fraction of water in the crystal solution or by changing the RH of the air surrounding a crystal. Many successful examples are in the literature (Adachi *et al.*, 2009; Bowler *et al.*, 2006; Cramer *et al.*, 2000; Fratini *et al.*, 1982; Gupta *et al.*, 2010; Heras *et al.*, 2003; Hu *et al.*, 2011; Kadlec *et al.*, 2011; Kuo *et al.*, 2003; Nakamura *et al.*, 2007; Sam *et al.*, 2006; Vijayalakshmi *et al.*, 2008; Yap *et al.*, 2007; Zerrad *et al.*, 2011). Several specific devices have been developed to control the humidity surrounding a crystal (Einstein, 1961; Sjogren *et al.*, 2002; Pickford *et al.*, 1993) with modern devices mounted at X-ray sources or synchrotron beamlines (Kiefersauer *et al.*, 2000; Russi *et al.*, 2011; Sanchez-Weatherby *et al.*, 2009). The ability to change the relative humidity while characterizing changes *via* diffraction allows any changes undergone by the crystal to be seen in real time and increases the chances of characterizing a beneficial phase change.

The HC1 was developed at the EMBL Grenoble to be a user friendly device compatible with a complex beamline environment (Sanchez-Weatherby *et al.*, 2009). It produces an air stream with a controlled relative humidity using a dispensing nozzle, in the same manner as cryo-stream devices produce a nitrogen flow at 100 K, and is therefore easy to integrate with most diffractometers. It supplies a stream of humid air at a RH determined by a dew point controller acting on a water saturated air supply. The device is now installed at laboratories and synchrotrons across the world (Bowler, *et al.*, 2015), resulting in many successful experiments (Hu *et al.*, 2011; Kadlec *et al.*, 2011; Malinauskaite *et al.*, 2014; Oliete *et al.*, 2013). The device can also be used for ambient temperature data collection (Bowler, *et al.*, 2015; Russi *et al.*, 2011) where the RH must be matched to the mother liquor to prevent dehydration of the crystal. The first step in these experiments is to define the equilibrium point between the RH and the mother liquor of the sample. This is an essential step as it defines the starting point for the experiments and maintains the crystal in a stable environment when the mother liquor is removed. In order to facilitate this stage we measured the equilibrium RH points for a variety of solutions commonly used for the crystallization of proteins and nucleic acids (Wheeler *et al.*, 2012). This provided a starting point for most experiments and the results obtained were compared with Raoult's law for the equilibrium vapour pressure of water above a solution (and for solutions of polymers, with a generalisation). The measurements made were consistently higher than those predicted by Raoult's law and a satisfactory explanation for the discrepancy could not be found. Here, we have repeated the measurements using a device based on the HC1 but with higher precision in the control of RH. The new measurements are in very good agreement with Raoult's law. Because of its importance, we present a simple explanation for Raoult's law using statistical mechanics and also show how this treatment can be extended to polymer solutions, where Raoult's law breaks down. These results illuminate the machinery underlying a long observed phenomenon and allow the accurate prediction of humid atmospheres for specific sample requirements that is applicable to a wide variety of fields.

## 2.0 Experimental Procedures

*Relative humidity measurements*

Solutions of PEG were made gravimetrically at concentrations between 50.0% and 10% w/w. Stock solutions of salts at 3 M were made and then diluted to reach the desired concentration. A round 600 μm Micromount (MiTeGen, Ithica, NY, USA) was mounted on either the BM14 or MASSIF-1 (Bowler, *et al.*, 2015; Nurizzo *et al.*, 2016) diffractometers with a HC-Lab (Arinax, 259, Rue du Rocher de Lorzier, 38430 Moirons, France) mounted at a distance of 5 mm from the loop. The HC-Lab is based on the original HC1 developed at the EMBL, Grenoble (Sanchez-Weatherby *et al.*, 2009) but with improvements in the dew point controller, temperature measurement, and calculation of relative humidity. These developments have led to a device with superior control and stability of relative humidity levels. In order to determine the equilibrium RH a drop of solution (typically 2 μl) was placed on the loop with a pipette. The diameter of the drop was measured using specific image analysis software. The humidity was adjusted until the drop diameter was stable. This was repeated a few times until the drop diameter was stable upon initial placement on the loop. Each measurement was repeated three times at ambient temperatures between 21 and 24.0$^{o}$C.

## 3.0 Results

*3.1 Agreement between measured equilibria and predicted values*

In previous work we measured the RH equilibrium points for a range of solutions commonly used in protein crystallization and examined the results in terms of Raoult's law and the Flory-Huggins model for the entropy of mixing of polymers (Bowler, *et al.*, 2015; Wheeler *et al.*, 2012). While the measured values provided a starting point for humidity control experiments and Raoult's law should be a good explanation for the observed results, there was a considerable discrepancy between the two (Wheeler *et al.*, 2012). Measured values were consistently 1 to 3% higher than those predicted which were attributed to the condenser used in the device being rather inaccurate at humidity values above 96%. Repeating these measurements using the new humidity control device, the HC-lab, the discrepancy is no longer significant (Figures 1A and 2A). The results now obtained from the HC-lab are also in agreement with detailed studies of the activity of water above salt (Robinson, 1945; Wishaw & Stokes, 1954) and polymer solutions (Sadeghi & Shahebrahimi, 2011; Sadeghi & Ziamajidi, 2006) (Figures 1B and 2B), with the salt solution measurements made in this study appearing to be more accurate. This now brings the control of relative humidity surrounding crystals into line with measurements made using dedicated and accurate devices as well as theoretical calculations.

*3.2 Derivation of the origin of Raoult's Law*

Raoult's Law is for the reduction of the saturated vapour pressure above a solvent when a mole fraction $x$ of some solute is dissolved within it. If the vapour pressure above the pure solvent is $p_0$ then the vapour pressure of the solvent above the solution is given by

$$p = p_0(1 - x) \qquad (1)$$

It is of course an idealisation, but is remarkably good, particularly at low mole fractions of the solute. Originally empirical (Raoult, 1887), from what principles can it be derived? Any such derivations depend on the assumption of an ideal solution, meaning that within the body of the solution the elements of the solute are near identical to the elements of the solvent (and yet for a non-volatile solute the solute cannot enter the vapour phase). In thermodynamics, equilibrium at constant temperature and pressure corresponds to a minimum of the Gibbs' function G and hence liquid-vapour equilibrium requires equal chemical potentials. The chemical potential of the solvent vapour phase is the same as that of the solvent, both above the pure liquid solvent and above a solution. The chemical potential in the solution is reduced by mixing; thermodynamic arguments are used to turn an entropy of mixing into the change of chemical potential. Thermodynamics does not deal with the mechanisms underlying these steps and it seems reasonable to ask, first, how the vapour pressure can be affected by the number of ways of arranging fixed numbers of two kinds of molecules and, secondly, why is there no apparent role for a work function related to the latent heat of vaporisation?

Raoult's Law is the direct result of the dilution of the solvent by the solute and can be extracted by applying elementary statistical mechanics. The machinery involves the energy levels the confined components can occupy and, in the simplest case of non-ideal solutions, differences in work functions are both important and easily calculated.

*3.2.1 Statistical mechanics*

It is a truth universally acknowledged that any system (such as an atom in a box) that has energy levels $\epsilon_i$ and is in thermal equilibrium at temperature *T* has a probability of occupying a given level proportional to $\exp(-\frac{\epsilon_i}{kT})$ where *k* is Boltzmann's constant. This is because the vast majority of possible configurations consistent with a prescribed total energy have this distribution. Thus for the macroscopic phenomena we are concerned with sums or averages over very many individual microscopic systems (here atoms, ions or molecules). For pure solvent we divide the energy levels into two classes, those in the liquid and those in the vapour phases. They are separated by a step in energy, a work function *W*, and so the number $n_i^v$, from a total of *N* atoms, found in a state of energy $\epsilon_i^v$ above *W* in the vapour is given by

$$n_i^v = N \exp(-\frac{[\epsilon_i^v+W]}{kT})/\{\sum_j \exp\left(-\frac{[\epsilon_j^v+W]}{kT}\right) + \sum_k \exp(-\frac{\epsilon_k^l}{kT})\} \qquad (2)$$

For a given temperature, the total number of atoms in the vapour is found by summing the numerator above over the index $i$. As the vapour energies start above the energy levels in the liquid by the work function $W$ (closely related to the latent heat) the fraction of atoms in the vapour contains a suppression factor of $\exp(-\frac{W}{kT})$. We are not yet concerned with this factor, or with the details of the structure of the energy levels. It suffices that for a given temperature and container, the number of atoms in the vapour phase is a fraction $y$ of the total number of solvent atoms $N$. The fraction $y$ is determined by the work function, the temperature and by the detailed structure of the energy levels, determined by the volumes available. If a fraction $x$ of the solvent atoms are removed and replaced by $Nx$ units of solute, changing nothing else, the volume of the container does not change and neither the detailed structure of the energy levels nor the work function for solvent atoms change because of the close identity of the solvent and solute units in an ideal solution. The fraction of solvent atoms in the vapour phase does not change and because there are now only $(1-x)N$ atoms of solvent, the number of atoms of solvent in the vapour phase is reduced by a factor $(1-x)$. Hence, the reduced vapour pressure and Raoult's law.

This simple argument is indubitably correct, given the assumptions of an ideal solution. The flux of solvent molecules leaving the surface is reduced by a factor $(1-x)$ and for equilibrium both the returning flux and the number density of solvent molecules in the vapour phase are also reduced by a factor $(1-x)$; directly the result of the smaller concentration of solvent molecules. This approach can be extended to non-ideal solutions (such as solutions of polymers) but is more complicated because of the need to calculate differences in work functions.

*3.2.2 Some technical details concerning volume*

A second result from elementary statistical mechanics removes a potential objection to the above argument. What if the volume of pure solvent is reduced? If the volumes of liquid and of vapour are being held constant, the number of vapour atoms is (for fixed temperature) a definite fraction of the number of atoms in the liquid phase. The more general result is that the concentration of atoms in the vapour phase is a definite fraction of the concentration of atoms in the liquid phase. The vapour pressure above a liquid in a sealed container does not, in equilibrium, depend on the volume of liquid in the container. Thus (1-$x$)$N$ atoms of solvent in the container without $xN$ atoms of dissolved solute, would not (and does not) result in a pressure reduced by (1-$x$). The reason is rather technical; the energy levels for atoms in the vapour are those of particle waves confined within the volume between the liquid surface and the walls of the container. For an ideal gas, the number of energy levels in a given interval of energy is proportional to the volume – the spacing goes down as the volume goes up. If the volume available to vapour doubles, the number of levels in some interval $\Delta\epsilon$ at $\epsilon$ also doubles and hence so does the number of molecules in the vapour. Thus the concentration of atoms in the vapour phase is constant as the volume increases - the pressure remains the same. Similarly, the molecules in the liquid roam throughout the liquid volume and their wave functions are constrained by the walls (and the liquid surface). If the volume of liquid is reduced, the sum over the populations of liquid energy levels is reduced because there are fewer of them. The spacing of energy levels in the liquid has gone up with the reduction in volume and the concentration in the liquid remains the same. Thus the saturated vapour pressure above the liquid remains constant as the ratio of vapour volume to liquid volume is increased, until of course all the atoms originally in the liquid are in the vapour phase. Thereafter, as the volume is increased (by pulling back on a piston perhaps) the vapour density and so the pressure along the isotherm fall.

When solvent molecules extracted are replaced by solute, the solute molecules make up the missing liquid volume. This makes available to the reduced number of solvent molecules the same energy level structures in both the liquid and vapour phases. This dependence of the energy level density on free range volume results in the concentration of atoms in the vapour phase being a definite fraction of the concentration in the solution. It is important for considering the vapour pressure above solutions that are not ideal; for example, polymers. Finally, it is essential for understanding the thermodynamic treatment and entropy of mixing.

*3.2.3. Solutions of molecules of different sizes*

Suppose now that instead of replacing a fraction of molecules of solvent with molecules of solute pre-empting the same volume, solute molecules require a different volume. For the case of polymers, such as polyethylene glycol, the specific volume will be larger, substantially larger for the heavier long chain polymers. Let there be $N_1$ molecules of solvent of specific volume $v_1$, similarly for the solute $N_2, v_2$. The volume occupied by the liquid solution is $N_1 v_1 + N_2 v_2$ and the concentration of solvent molecules is less than for pure solvent occupying the same volume. The ratio of concentrations of the solvent molecules in the solution to pure solvent gives a factor in the vapour pressure ratio of

$$\frac{N_1 v_1}{N_1 v_1 + N_2 v_2} \qquad (3)$$

This factor reduces to Raoult's law as the specific volumes of solvent $v_1$ and solute $v_2$ approach equality. This is not the whole story because the difference in work functions for solvent phase transitions is not zero except in this limit. The following simple calculation yields the requisite difference in work functions. The work function is the work that has to be done when removing a molecule against the cohesive forces in the liquid and any contribution from ambient pressure. If a molecule is instead added, the volume it pre-empts acquires negative potential energy and hence the work done is - $P\Delta V$. Consider the operation of replacing a molecule of solvent by one of solute. The liquid expands by a volume $\Delta V = (v_2 - v_1)$ and this volume contains a negative potential energy density. The effective pressure P must balance that from the thermal energy density and so is given by

$$P(N_1 v_1 + N_2 v_2) = (N_1 + N_2)kT \qquad (4)$$

Thus, the work that has to be done to make the replacement is given by

$$-\frac{kT(N_1 + N_2)(v_2 - v_1)}{N_1 v_1 + N_2 v_2} \qquad (5)$$

This is made up of two pieces, the work necessary to insert a molecule of solute (a contribution to the chemical potential $\mu_2$) and the work necessary to extract a molecule of solvent ($-\mu_1$). As $N_2 \to 0$, the solution approaches pure solvent, identifying the difference in the work that has to be done to deliver one molecule of solvent to the solution as opposed to pure solvent as

$$\Delta W^\uparrow = \frac{kT N_2 (v_2 - v_1)}{N_1 v_1 + N_2 v_2} \qquad (6)$$

This can be verified by calculating the work done against pressure to insert a solvent molecule into the solution as opposed to the same volume of pure solvent. Calculate the (pressure related) work done inserting an atom of 1 into a solution and also calculate the work done inserting an additional atom into a volume of pure species 1. In both cases $\Delta v = v_1$. The pressure in solution is given by

$$\frac{(N_1 + N_2)kT}{N_1 v_1 + N_2 v_2}$$

as in (*eq. 4*) and the pressure in pure solvent is $kT/v_1$. Then

$$-\Delta(p\Delta v) = -\{\frac{(N_1 + N_2)v_1}{N_1 v_1 + N_2 v_2} - 1\}kT$$

and this also yields (*eq. 6*).

The difference in work functions for removing atoms to the vapour phase, $\Delta W^\downarrow$, is the negative of (*eq. 6*). The effect on relative humidity is an exponential factor

$$\exp\left(-\frac{\Delta W^\downarrow}{kT}\right) = \exp(\frac{\Delta W^\uparrow}{kT})$$

The concentration ratio *(eq. 3)* multiplied by this factor yields the relative humidity of the solvent

$$\frac{p}{p_0} = \frac{N_1 v_1}{N_1 v_1 + N_2 v_2} \exp\{\frac{N_2(v_2 - v_1)}{N_1 v_1 + N_2 v_2}\}$$

*(7)*

The first factor on the right hand side is the volume fraction of solvent in the solution and reduces to Raoult's law as the specific volumes become equal. The second factor goes to unity in this same limit. It is less obvious that (*eq. 7*) also reduces to Raoult's law in the limit of extreme dilution, regardless of the ratio of specific volumes – but it is so.

This expression (*eq. 7*), derived using elementary notions from statistical mechanics, is the same as that derived using thermodynamics and the Flory-Huggins entropy of mixing devised for polymer solutions (Flory, 1942; Flory, 1970) or, equivalently, Hildebrand's entropy of solution of molecules of different size (Hildebrand, 1947). In such treatments both factors in (*eq. 7*) emerge from matching chemical potentials. Our treatment clarifies the physical meaning of the factors – the first factor is concentration ratio; the second (exponential) factor embodies the difference in work functions. In the appendix we discuss the relationship between simple statistical mechanics and thermodynamic arguments, addressing in particular the significance of entropy of mixing.

**4.0 Discussion**

The control of the relative humidity surrounding samples is important to maintain their integrity and study the effects of increased or decreased humidity. Here we have established that the theoretical calculation of RH values we previously determined (Bowler, *et al.*, 2015; Wheeler *et al.*, 2012) are in satisfactory agreement with a humidity control device used on protein crystallography beamlines. As the predicted values are also in complete agreement with measurements made using specific devices, the previous discrepancies can be ascribed to shortcomings in the control of RH in the HC1c used. We have also determined the origin of the observed vapour pressure changes above solutions of solutes. If N units of a liquid solvent are in an equilibrium where liquid and vapour phases coexist, a fixed fraction are (for a given temperature) in the vapour phase. If the number of units is reduced to N(1-x), if all else remains unchanged, because of the presence of Nx units of the solute in an ideal solution, then the number of units in the vapour phase (and hence the pressure) are reduced by the same factor (1-x), Raoult's Law. For unequal sizes of solvent and solute components, the dilution factor has to be multiplied by an exponentiated work function. These results provide a solid basis on which to predict the relative humidities required to maintain a wide variety of samples and solutions in homeostasis.

**Acknowledgements**

M.W.B thanks Babu Manjestay and Hassan Belrahli (EMBL, Grenoble) for access to the HC-Lab on beamline BM14 and Ralf Siebrecht and Jaouhar Nasri (Arinax) for stimulating discussions.

**Figure legends**

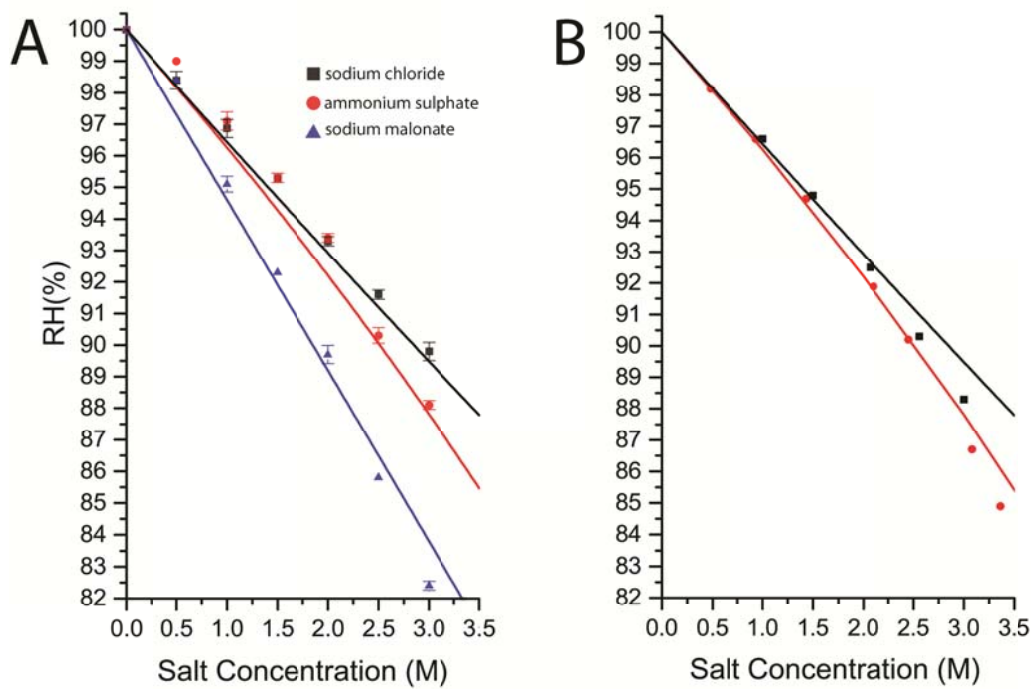

**Figure 1.** Plot showing the equilibrium relative humidity for salt solutions commonly used as precipitants or additives in macromolecular crystallogenesis measured using the HC-Lab (**A**) and the measured vapour pressures above solutions of ammonium sulphate (Wishaw & Stokes, 1954) and sodium chloride (Robinson, 1945) solutions (**B**). The lines show the calculated RH from Raoult's law (Wheeler *et al.*, 2012). The measurements made using the HC-Lab (**A**) more accurately reflect the predicted values from Raoult's Law.

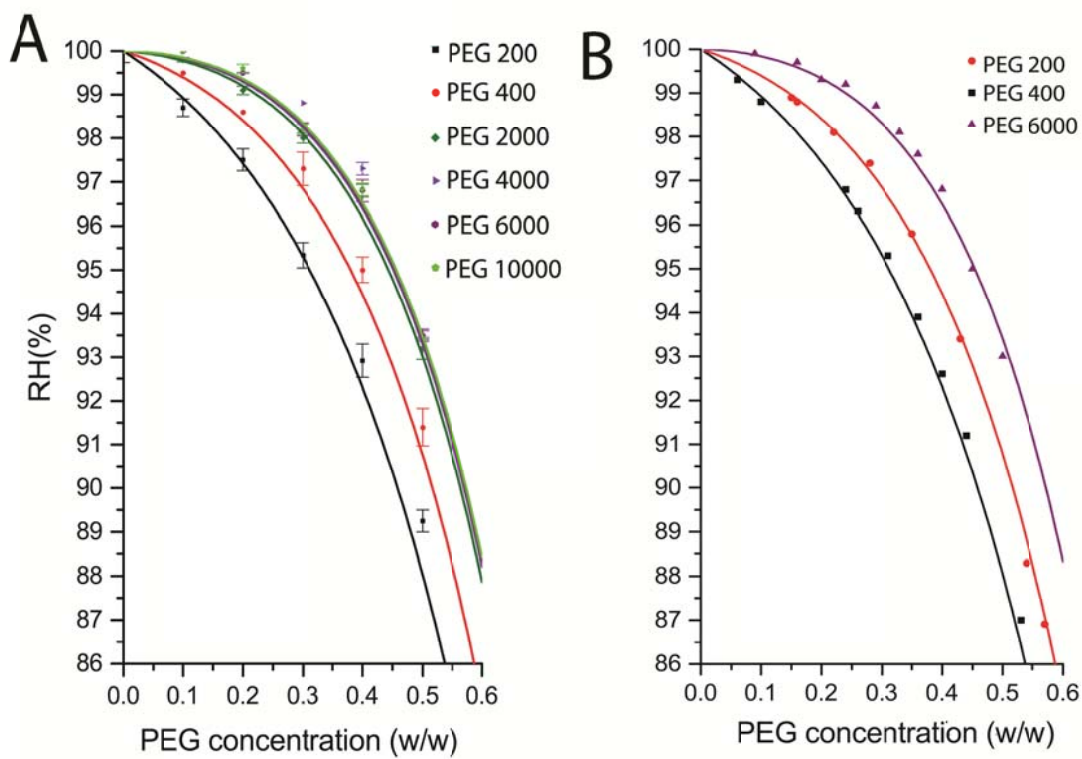

**Figure 2.** Plot showing the equilibrium relative humidity for PEG solutions commonly used as precipitants or additives in macromolecular crystallogenesis measured using the HC-Lab (**A**) and the measured vapour pressures above PEG solutions from (Sadeghi & Shahebrahimi, 2011; Sadeghi & Ziamajidi, 2006) (**B**). The lines show the calculated RH from Raoult's law modified for polymer solutions (Bowler, *et al.*, 2015).

**Appendix. Statistical mechanics, entropy and chemical potentials**

The origin of Raoult's law lies in the freedom of the units of solvent and of solute to roam all through the volume of liquid. For the assumptions of an ideal solution access of both solvent and solute to the whole volume results in energy levels available (to the solvent) for a given volume, unchanged from those in pure solvent and the density of (energy) states is proportional to volume. Entropy of mixing codes these same ideas in the language of thermodynamics.

Suppose that we can decompose a system into many identical pieces having energy levels $\epsilon_i$. This complex system is in thermal equilibrium at some temperature $T$, with an exponential distribution in energy of the components. Let the total energy of our complex system be $U$. The following relations apply:

$$n_i = \frac{N \exp(-\frac{\epsilon_i}{kT})}{z}$$

$$U = \sum_i n_i \epsilon_i = N \frac{\sum_i \epsilon_i \exp(-\frac{\epsilon_i}{kT})}{z}$$

where
$$z = \sum_i \exp(-\epsilon_i / kT)$$

A small change in the internal energy $U$ can be written as

$$\Delta U = \sum_i \epsilon_i \Delta n_i + \sum_i n_i \Delta \epsilon_i + \mu \sum_i \Delta n_i$$

The last term corresponds to taking a common potential out from the levels $\epsilon_i$ and vanishes if there is no change in the number of components in the system. The first term is the result of slightly redistributing the population over the energy levels $\epsilon_i$. It represents the addition of heat. The second term corresponds to the energy levels changing with no change of population – doing it very slowly. If the volume slowly increases the energy levels slowly sag and the system does work. Thus the equivalent expression in thermodynamics is the first law in the form

$$\Delta U = \Delta Q + \Delta W + \mu \Delta N$$

where the last term is called chemical work and $\mu$ is the chemical potential. (Generally, each species of atom has its own chemical potential.) If everything is done very slowly and reversibly

$$\Delta Q = T\Delta S$$

where $S$ is the thermodynamic entropy going back to Carnot. We are identifying the heat term in the first law with

$$\Delta Q = \sum_i \epsilon_i \Delta n_i$$

Express the energy of the $i$th level in terms of its population

$$\epsilon_i = -kT ln(P_i z) \qquad P_i = n_i/N$$

For fixed $N$ the sum of $\Delta n_i$ is zero and so the term in $\ln(z)$ drops out and

$$\Delta Q = -kTN\Delta\{\sum_i P_i ln P_i\} = T\Delta S$$

demonstrating the equivalence of the Carnot and Boltzmann entropies.

Thus the entropy associated with $N$ units (atoms, molecules, ions …) distributed over these energy levels, in equilibrium at temperature $T$, is

$$S = -Nk \sum_i P_i ln P_i$$

The probabilities $P_i$ involve the normalising factor $z$, a sum over all energy levels. The value depends on the level density. The more levels the components are spread over, the smaller the individual $P_i$ and the larger the entropy.

Substitution of the expressions for the exponential probabilities yields for the entropy

$$S = \frac{E}{T} + Nk\ln z$$

Rewriting the sum for *z* as an integral

$$z = \int \exp(-\epsilon/kT) \frac{dn}{d\epsilon} d\epsilon$$

The density of states factor $\frac{dn}{d\epsilon}$ depends on $\epsilon$ (which integrates out) and is linearly proportional to the volume available to the wandering molecules. Consider taking a volume $V_1$ of solvent and a volume $V_2$ of solute. (This is most easily envisaged if the solute is also a liquid. Otherwise pretend that there is a solute liquid with the properties the solute will display in the solution). Before mixing the two together each has its entropy, appropriate to volumes $V_1$ and $V_2$ respectively. After mixing, both solvent and solute have access to a total volume $V_1 + V_2$. For an ideal solution nothing else has changed and so taking the difference of entropy after and before the mixing

$$S_{12} - S_1 - S_2 = k\{(N_1 + N_2)\ln(V_1 + V_2) - N_1 \ln V_1 - N_2 \ln V_2\}$$
(A.1)

This is the entropy of mixing and arises entirely from the increased density of energy levels as more volume is made available for units of both solvent and solute to roam at random. (These volumes are defined by the boundaries confining the liquids, setting boundary conditions and hence determining the quantised energy levels.)

Since we are looking at mixing of two forms of condensed matter, each with the same specific volume (ideal solution again) the above expression for the entropy of mixing can also be written with $V_1$ and $V_2$ replaced by $N_1$ and $N_2$ respectively. The result is essentially identical to the product of *k* (Boltzmann's constant) and the logarithm of the number of different ways of arranging $N_1$ and $N_2$ units. This is a purely combinatorial problem and the number of perceptibly different ways is given by

$$\frac{(N_1 + N_2)!}{N_1! N_2!}$$

What can this have to do with the vapour pressure above a solution? We now see that $N_1$ and $N_2$ are (for an ideal solution) proxies for $V_1$ and $V_2$ and these volumes control the energy levels available to the components of the solution before and after mixing.

More generally, suppose that the solvent molecules are associated each with free volume $v_1$ and the solute molecules with $v_2$. Then the entropy of mixing (A.1) above is

$$k\{(N_1 + N_2)\ln(N_1 v_1 + N_2 v_2) - N_1 \ln(N_1 v_1) - N_2 \ln(N_2 v_2)\} \quad (A.2)$$

This is essentially the expression for the entropy of mixing for solvent and solute molecules of different free volumes to be found in *(eq. 3)* of Hildebrand (1947), where the volumes are introduced through a classical argument concerning uncertainty of location. This is also equivalent to the Flory-Huggins entropy for polymer solutions, most clearly discussed in Flory (1970).

The derivative of the entropy of mixing with respect to the number of solvent molecules within the solution ($N_1$) yields the difference in chemical potentials that must match the difference in chemical potentials of the vapours above solution and pure solvent.

$$\Delta\mu = -T\frac{\partial \Delta S}{\partial N_1}$$

For the solution

$$\mu_{12}^1 = -kT\frac{\partial}{\partial N_1}(N_1 + N_2)\ln(N_1 v_1 + N_2 v_2)$$
(A.4)

and for the pure solvent before mixing

$$\mu_1^1 = -kT\frac{\partial}{\partial N_1}N_1\ln(N_1 v_1)$$
(A.5)

Taking the difference and matching to the vapour phase eventually yields

$$\ln\left(\frac{p}{p_0}\right) = \ln\frac{N_1 v_1}{N_1 v_1 + N_2 v_2} + \frac{N_2(v_2 - v_1)}{N_1 v_1 + N_2 v_2}$$

equation (6) of Hildebrand (1947). Then the relative humidity of the solvent above such a solution is

$$\frac{p}{p_0} = \frac{N_1 v_1}{N_1 v_1 + N_2 v_2} \exp\{\frac{N_2(v_2 - v_1)}{N_1 v_1 + N_2 v_2}\}$$
(A.6)

Equation (A.6) is identical to (*eq. 7*).

In section 3.2.3 we calculated the difference in work functions for the solvent in a solution of volume V and pure solvent in the same volume. This result can also be obtained from the differential of the difference of entropies of the solution and pure solvent in equal volumes. The only terms that survive in the difference are $(N_1 + N_2)k \ln V$ for the solution and $N_1^0 k \ln V$ for the pure solvent. The volume $V$ is

$$V = N_1 v_1 + N_2 v_2 = N_1^0 v_1$$

The relevant difference in chemical potentials is then

$$kT \frac{N_2(v_2 - v_1)}{N_1 v_1 + N_2 v_2}$$
(A.7)

The negative of this is the difference in work functions, needed to complete the ratio of vapour pressures at the end of section 3.2.3. The result (A.7) above of course agrees with (*eq. 6*).